# Architecture Definition in Complex System Design Using Model Theory

C. Dickerson[1], *Senior Member, IEEE*, M. Wilkinson[2], E. Hunsicker[3], S. Ji[4], *Member, IEEE*, M. Li[5], *Member, IEEE*, Y. Bernard[6], G. Bleakley[7], and P. Denno[8]

*Abstract*—Architecture Definition, which is central to system design, is one of the two most used technical processes in the practice of model-based systems engineering. In this paper a fundamental approach to architecture definition is presented and demonstrated. The success of its application to engineering problems depends on a precise but practical definition of the term architecture. In the standard for Architecture Description, ISO/IEC/IEEE 42010:2011, a definition was adopted that has been subsumed into later standards. In 2018 the working group JTC1/SC7/WG42 on System Architecture began a review of the adopted definition, holding sessions late in the year. This paper extends and complements a position paper submitted during the meetings, in which Tarski model theory and ISO/IEC 24707:2018 (logic-based languages) were used to better understand relationships between system models and concepts related to architecture. Independent from the working group, it now contributes intuitive fundamental definitions of the terms architecture and system that are used to specify a mathematically based technical process for architecture definition. The engineering utility and benefits to complex system design are demonstrated in a diesel engine emissions reduction case study.

*Index Terms*— Architecture, system, definition process, model theory, Category Theory, diesel emissions

## I. Introduction

ARCHITECTURE is key to the modern practice of engineering but in many ways, a precise practical definition has been elusive if not ineffable. The term would be understood by a general audience as a property of buildings or large-scale structures. Although understanding architecture in this way is intuitive and useful, it lacks the precision needed for application to engineering problems. The position taken in this paper is the same as the one taken with JTC1/SC7/WG42 in late 2018: a prose definition of a technical term should be complemented by a mathematical interpretation [1]. This is a model theoretic method for assessing validity of definitions that is essential for specifying an architecture definition technical process for engineering and scientific problem solving.

### A. Ubiquity of the Architecture Metaphor

In civil engineering, architecture relates a building's purpose (function), form (how its spaces are organized to achieve the purpose), and construction (what it is built from and how it is built). Beyond the design and construction of buildings, architecture has found a place in many other different disciplines, serving as a readily apprehended metaphor by which the native 'structures' of a subject can be understood.

Although architectural ideas are now prevalent in disciplines as diverse as civil engineering, systems engineering, management science, biology and mathematics; there is no consensus on terminology or meaning – there is common ground but there is no unity. The potential of achieving precise unified definitions by means of a formal approach, and the benefits of so doing have been long recognized in, for example: (i) the foundational work of Bertalanffy [2], which was inclusive of many mathematical expressions of systems concepts; (ii) Wymore's codification of model-based systems engineering (MBSE) [3], which expressed a programme for systems engineering; and (iii) Rosen [4], who was perhaps the first to recognize the possibility of using Category Theory for systems and scientific problem solving. The authors have also long recognized and investigated formal approaches in: (i) using architecture to develop and acquire mission level capabilities [5]; (ii) interpretation of an adopted definition of 'system' as a Hamiltonian system in physics [6]; and (iii) more recently the Wilkinson polemic paper [7]. In a shift from a narrow systems orientation, recent efforts within JTC1/SC7/WG42 have applied the term architecture to entities not normally considered to be systems. Note however, that the purpose of this paper is neither to report on such work nor to critique it. Rather the purpose is to offer definitions of architecture and system which are complemented by mathematical interpretation that can be: (i) used for specification of a technical process for architecture definition; and (ii) exploited in engineering practice and in relevant standards.

Corresponding author: C. Dickerson (c.dickerson@lboro.ac.uk).
[1]Wolfson School of Mechanical, Electrical and Manufacturing Engineering, Loughborough University, Loughborough, LE11 3TU, UK
[2]BAE Systems Maritime, Barrow-in-Furness, Cumbria, LA14 1AF, UK
[3]School of Science, Loughborough University, Loughborough, LE11 3TU, UK
[4]Department of Computer Science, University of York, York, YO10 5GH, UK
[5]Rolls-Royce plc. Victory House, Victory Road, Derby, DE24 8BJ
[6]Airbus Defence and Space SAS, 31 Avenue des Cosmonautes, 31400 Toulouse, France
[7]IBM UK Ltd, North Harbour, Portsmouth, UK
[8]National Institute of Standards and Technology, 100 Bureau Drive, Gaithersburg, Maryland, USA

*B. Precise Natural Language Definitions*

The ambiguities introduced by using natural language in the definition of technical terms can be reduced and possibly resolved by introducing languages that are more precise or formal; the predicate calculus of logic being just one. An issue though is that such languages may not be accessible to a general audience or to the stakeholders in a discipline.

A pioneer of digital computer architecture, F.P. Brooks [8], proposed a resolution to this issue that remains valid today,

*One needs both a formal definition of a design, for precision, and a prose definition for comprehensibility.*

This idea was adopted as a 'principle of definition' by Dickerson and Mavris [9] and will be adapted to reason about the terminology of architecture and systems. The term '*a formal definition of a design*' will be replaced by '*a mathematical interpretation of a concept*'. Concordance between the two types of definition is achieved by interpreting natural language (prose) into mathematical logic and models. The primary challenge in this approach is to preserve concordance between precision and comprehensibility without losing either.

The position paper [1] put forward to the working group JTC1/SC7/WG42 applied this principle to the definition of (System) Architecture in [10]. This definition will be referred to as the 'adopted' definition. As previously mentioned, recent work within WG42 has generalized the adopted definition to refer to entities other than systems.

It was argued in the position paper that there is a need for distinction between key terms such as: *concept, property, embodiment, element, and relation*. Based on Tarski model theory [11] [12], an initial attempt to refine the wording of the adopted definition was made to illustrate how a better distinction could be achieved. Terminology for *object, property*, *class, and type* from mathematics will now also be used for further precision and clarity. For example, *class* is a technical term in set theory for defining a collection of objects that share one or more common properties.

Fundamental definitions are now offered independent of the working group that can be used to make a distinction between key terms and express an intimate relation between architecture and structure. The first definition proposed is:

*Structure* is junction and separation of the objects of a collection defined by a property of the collection or its objects.

This definition is at a higher level of abstraction than that of the adopted definition of architecture, but it is readily applicable to physical examples. In civil engineering for example, a building is a collection of objects that includes rooms, which are joined as well as separated to achieve a defined purpose. Buildings in civil engineering are referred to as *structures*.

One mathematical interpretation of the proposed definition of structure is the *separation* of the collection of the counting numbers $\mathbb{N}$ into two disjoint subsets based on the *property* of divisibility by the number 2. One of the subsets is comprised of the even numbers, and the other the odd numbers. The *junction* (union) of the two subsets comprises the whole of $\mathbb{N}$. This type of structure in mathematics is called a partition. It separates the underlying set into equivalence classes. Partitions of a collection into equivalence classes are common in abstract algebra; and more generally in Category Theory [13]. Thus, structure as defined in this paper adheres to the principle of definition.

The mathematical term *object* is used in the definition because it is an abstraction that is only constrained by the properties it possesses, e.g. functional, physical, or temporal. Note that *elements* (of a set) are mathematical objects. 'Elements' is a usage of terms in alignment with the adopted definition of architecture. However, it should be noted that although the term structure is used widely in the engineering community, there is no common definition offered in current relevant standards.

In specific domains such as software engineering, a term such as *stable binding* is often used instead of junction. Thus, the proposed definition of structure accommodates ideas such as stable bindings of static objects, instantaneous bindings (events), as well as dynamic behavior. It is also worth noting that most definitions of structure only mention some form of junction; but they are silent on separation. However, these two terms should be on an equal footing. For example, when a system boundary is defined, a collection of objects of interest is *separated* into members of a system and those of an environment.

Structure therefore expresses a relation among the objects of a collection. The general character of this relation is a specialized property which is referred to as a *type*. In software engineering, the term *classifier* is often used instead of type. This leads to the second definition:

*Architecture* is structural type in conjunction with consistent properties that can be implemented in a structure of that type.

Such properties will be called *architectural properties*. Note then that every architecture is associated with at least one architectural property; its structural type. Architectural properties can constrain and further specify structure. An example was investigated in detail in the position paper [1]. In engineering practice, architecture can be considered as a qualified structural type, i.e. one with qualities (e.g. implementable properties) that can achieve a defined purpose.

The association of a structural type with a collection of properties in the proposed definition of *architecture* can then be represented as an ordered pair. This construct can also provide a formal underpinning to architecture frameworks, which are commonly visualized as a matrix. Each row can be defined by one or more architectural properties and each column can be a structural type, with the intersection (a cell in the matrix) defining a class of architecture implemented by the architectural properties in a way that is consistent with the structural type. An example is when the functionality of a system is implemented in the graphical structure of a use case in object-oriented modeling.

The fundamental definitions offered in this paper will be referred to as 'essential definitions'. They are subsistent in the sense of being an economical choice of generally understandable words that have mathematical interpretation. The essential definition of architecture is complemented by:

A *System* is a set of interrelated elements that comprise a whole, together with an environment.

This definition is adapted from a mathematically based one cited by Bertalanffy [2] who has used the term *interrelated*, which involves *relations among relations*. Also, the environment in his definition is on an equal footing with the whole (set of elements); but its relation to the elements is not specified. It is not necessary for the set as a whole to belong to the environment or even for it to interact with it. The definition though does not consider a system without making reference to an environment.

When a structure of interrelations has been identified or defined, the set is said to be endowed with an interrelational structure. (Note: mathematicians use the term *endowed* for the pairing of a set with a relational structure.) Thus, a system is a set of elements endowed with a structure that is of an interrelational type. In this sense, system is a realization of architecture. It will also be seen that the interrelational structure of a system can be used to specify architecture. This provides a mathematical basis for a process that will be called Essential Architecture Definition. The term is used in the sense of Yourdon structured analysis in which 'essential' means necessary or essence.

The essential definitions in this paper underlie those specified by ISO in [10] and [14]. Note, for example, that the concept of 'a set of interrelated elements' includes that of 'a combination of interacting elements.' There are numerous other definitions in relevant standards. Most consider the commercial and life-cycle aspects of engineering such as definition and management of the system configuration, translation of system definition into work breakdown structures, among others. The essential definitions are intended to complement the standards; not to be a replacement.

*C. Structure of the Paper*

The paper is structured around three key contributions: the essential definitions of the terms architecture and system, their interpretation into a mathematically specified technical process for architecture definition, and a detailed demonstration of engineering utility and benefits to complex system design. The Introduction has established the problem to be addressed and the context; and offered 'essential definitions'. Sections II and III provide an historical and theoretical background. Section II is a brief history of relevant definitions of architecture and significant points about their evolution. It also has a brief summary of early attempts at system architecting (which is another name for architecture definition), its relation to system engineering, and issues associated with the definition of terms. Section III provides an explanation of conceptual structures and Tarski model theory that should be accessible to a general audience. It concludes with a comparative analysis of the adopted definition of architecture and the essential definition. These will be seen to be in general agreement. Some ambiguities in the adopted definition will also be addressed.

Section IV builds on these foundations to show how an architecture definition process that supports system design can be implemented. This provides a demonstration of the validity of the essential definitions offered in Section I-B. Section V applies the process to the specification of a calibration system for diesel engine emissions reduction. Constraint driven design methods are applied to the calibration problem. This demonstrates the engineering utility of the proposed architecture definition process. Section V concludes with how the process has been used in an MBSE standard concerned with constraint driven design.

## II. Brief History of Architecture

It is instructive to understand how the definitions of architecture and related terms in systems engineering have evolved over time and how they have been influenced by other disciplines. Although this might seem to be only of theoretical interest, the incorporation of such definitions within key standards has a significant impact on how systems engineering is conducted in practice. This section begins with a brief summary of the diversity of meanings of systems architecting and moves on to describe how key definitions have evolved in canonical standards. The Model Driven Architecture (MDA) in object-oriented software standards provides an initial example of how the essential definition can be used in an explanatory way by refining the usage of the term architecture in software development. It concludes with a short section on architecture styles and patterns.

*A. System Architecting*

A detailed examination of the nature and impact of different meanings of system architecting was conducted by Emes et al in 2012 [15]. This work reported on research to capture and analyze beliefs about the meaning of architecting and related terms. It considered definitions from three authoritative sources, including ISO/IEC/IEEE 42010:2011 [10]; and from a set of interviews with practitioners. Twelve contentious questions about architecting were investigated and six different perspectives on the meaning of system architecting and its relationship to system engineering were described.

The most emphatic conclusion from the research was that a consensus on the meaning of architecture did not exist and there were many diverse interpretations. The value of explicit formal definitions, such as those found in standards documents, was therefore judged to be particularly significant. The paper noted that the lack of robust definitions for key terms made the development of standards more difficult. It also noted that its scope excluded systems of systems and enterprises, for which further diverse interpretations of key terms might be expected.

More recently, ISO/IEC/IEEE 42020 [16] has defined a standardized approach to architecting, including processes for architecture conceptualization and elaboration. ISO/IEC/IEEE 42030 [17] has defined a standard for Architecture Evaluation. It is too early to judge what impact these new standards will have but, over time, they are likely to influence understanding of architecting and the definitions in other standards. These latest standards make use of the same definition of architecture, as described in Section II-B and discussed in Section III-D.

*B. System Architecture*

The earliest attempts to standardize ideas about the meaning of architecture within systems engineering were in connection with the lifecycle processes standard, ISO/IEC 15288. The first published version (2002) was influenced by legacy software engineering standards, which conditioned the way in which architecture was conceived and also combined it with design into a single process called Architectural Design. This was separated into two processes in the 2015 update [14].

The software discipline had additional influences. In the early 1990s, the IEEE considered architecture to be "the organizational structure of a system or component". This provided a formulation for architecture as a system property, described in terms of static structure (elements and their relationships). In 1996, with the aim of using an architectural metaphor as a foundation for systems engineering, Hilliard et al [18] defined architecture as "the highest level conception of a system in its environment". This has several important ideas: architecture is 'high level', includes an 'external focus' and is a 'conception' in the imagination.

The ideas described by Hilliard et al were carried over into the definitions used in IEEE 1471:2000 [19], which defined architecture as "the fundamental organization of a system embodied in its components, their relationships to each other, and to the environment and the principles guiding its design and evolution." The same wording was used in the 2008 issue of ISO/IEC/IEEE 15288. The idea that architecture is subjective is fully embraced in these standards, which introduce architectural descriptions (i.e. models of architecture) as the vehicle for conveying information about the subjective conception of architecture. However, the role of models of architecture for communication has caused a widespread perception that architecture and model are equivalent: [20] for example, considers architecture as a graphical model (or representation).

In the latest published versions of the key standards, specifically ISO/IEC/IEEE 42010:2011 [10] and 15288:2015 [14], [1] the adopted definition of architecture refers to "fundamental concepts or properties of a system in its environment embodied in its elements, relationships, and in the principles of its design and evolution". Here, the earlier term "organization" has been replaced by "concepts or properties". Rather than organization being embodied in a static structure (elements and their relations), it is the concepts and properties that are embodied.

A slightly refined definition has been adopted in [16] and [17], in which the idea of embodiment is now referred to only in a note. In a further refinement, these standards associate architecture with 'architecture entity' rather than 'system'. Finally, echoing many of the points above, the practitioner's view of architecture, as expressed in the Systems Engineering Book of Knowledge [21], also associates architecture with abstraction, conceptualization and high-level structure.

*C. Model Driven Architecture*

Over the same two-decade time period, the Object Management Group™ (OMG™) has used architecture in standards for software development. For example, the Model Driven Architecture® (MDA®) is an approach to software design, development and implementation. MDA provides guidelines for structuring software specifications that are expressed as models [22]. In MDA the software architecture of an application is the basis for deploying the computer code.

MDA development focuses first on the functionality and behavior of a system by means of a platform-independent model (PIM) that separates business and application logic from underlying platform technology. In this way, functionality and behavior are modeled once and only once. The PIM is implemented by one or more platform-specific models (PSMs) and sets of interface definitions until the system implementation is complete. A PSM combines the specifications in the PIM with details about how a system *uses* a particular platform type; but not the details necessary to implement the system. In the separation of platform technology from functionality and behavior, MDA conforms to the precepts of structured analysis as put forth by Yourdon.

In MDA the models of functionality and behavior in the PIM are transformed into PSM models by associating new concepts not found in the PIM. The models of the PIM are thus joined by transformation with those of the PSM. This collection of models therefore has both junction and separation that express a defining property of the collection, i.e. an approach for software design, development, and implementation. This is the definition of *structure*. Furthermore, because the PIM and PSM are defined as *types* of models, this structure derives from a structural type. Consequently, in terms of the essential definitions of this paper, MDA can then be considered as an architecture vis-à-vis an approach.

*D. Architecture Levels, Styles and Patterns*

It is common within systems engineering practice guidelines and textbooks to identify an iterative or 'fractal' process structure, including architecture definition, at each of the process steps. From this perspective, architecture is developed at each level in the hierarchy of system levels of detail. Within the well-known NASA Systems Engineering Handbook [23], for example, the systems engineering process at each level includes five distinct steps (four definition processes and one implementation process) defining (i) context, (ii) requirements, (iii) architecture, (iv) and design; then (v) proceed to implementation if the lowest level of detail is reached, otherwise return to step (i). This kind of thinking can be made harmonious with the standards definitions, as long as the 'entity' at each level being architected is considered as a system in its own right.

Finally, architectural styles and patterns should also be mentioned. Like architecture itself, these are problematic terms with no universally accepted definitions. Most authors seem to use 'style' as shorthand to codify a correspondence between different architectures; whereas 'pattern' is a re-usable style that addresses a defined type of problem.

III. USING TARSKI MODEL THEORY

A general audience would understand the term 'model' as a representation of something or an excellent example [24]. In order to avoid confusion between models and architecture, a well-established definition of 'model' from mathematics and logic is reviewed and used to complement natural language definitions. In mathematics, diagrams or symbolic expressions are generally used to *represent* models. In this section the term 'concept', which is central to the definitions of architecture reviewed in the previous section, will be understood in terms of conceptual structures that can become models when interpreted mathematically.

---

[1] Permission to reproduce extracts from British Standards is granted by BSI Standards Limited (BSI). No other use of this material is permitted. (For the convenience of readers, the equivalent international reference is cited.)

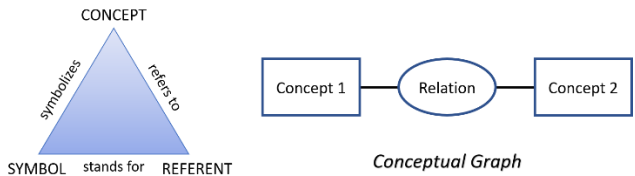

Fig. 1. Meaning Triangle and Concepts

*A. Motivation from Algebra*

The concept of orthogonality (right angles) in the plane geometry of triangles can be represented algebraically. If *a, b,* and *c* are the lengths of the sides of a triangle, and if these lengths have the algebraic relation, $a^2 + b^2 = c^2$; then the triangle is said to be a right triangle. Additionally, the two sides corresponding to $a$ and $b$ are said to form a right angle. This relation is an algebraic sentence that expresses a concept of orthogonality. Model theory is simply an abstraction of the idea that models are sentences interpreted into relational structures. The sentences can express theories or concepts.

*B. From Algebra to Logic: Model Theory*

The Propositional Calculus and the Predicate Calculus are languages of mathematical logic. Propositions are declarative statements represented by variables that take on the value of either true or false. This language of logic is inadequate for a proper theory of models.

Predicates, on the other hand, are declarative statements of relationships between variables and are represented by predicate letters $\{P_1, P_2 \ldots\}$ that represent relations among individual variables $\{v_1, v_2 \ldots\}$ as well as constants $\{c_1, c_2 \ldots\}$. A *sentence* in the Predicate Calculus is defined to be a fully quantified well-formed formula. The Predicate Calculus (with equality) is the level of precision in language necessary for a mathematical theory of models.

Tarski model theory offers the following simple but formal definition: a model is a relational structure for which the interpretation of a sentence in the Predicate Calculus becomes valid (true). This is called a *fully interpreted first order model; and is the basis for so-called model theoretic truth*. A relational structure in set theory is a set *M* and a collection of relations $\{R_\alpha\}$ on $M$.

Thus, the algebraic expression of orthogonality in the Tarski theory corresponds to the predicate sentence given by

$$\forall v_1, v_2, v_3 : P_1(v_1, v_2) = P_2(v_3) \quad (1)$$

Let the set *M* be the Cartesian plane. The interpretation of the predicates into relations assigns $P_1(v_1, v_2)$ to $a^2 + b^2$; and $P_2(v_3)$ to $c^2$, for specific values of *a, b,* and *c* that are lengths of the sides of a triangle (e.g. 3, 4, and 5). Every triangle ∆ in the Cartesian plane has a relation R∆ among its sides. When the relation conforms to $a^2 + b^2 = c^2$ then the predicate sentence is valid, and the triangle is a model of orthogonality. A more detailed mathematical review of the Predicate Calculus (with equality) and Tarski model theory can be found in [25]. A general but technical review for the non-specialist is given by the model theorist Wilfrid Hodges in the Stanford Encyclopedia of Philosophy [26].

*C. Comparison with Conceptual Structures*

The framework of Conceptual Structures described by Sowa [27] is essentially an approach to representing knowledge inside computer systems; but it is more general and can be applied more widely. In this framework a generic concept is a mental interpretation of a percept (which is a correlation of sensory impressions of the physical world). In this case the concept is called concrete. Concepts that have no such percepts are called abstract. Sowa points out that although concrete concepts may have direct associations with sensory impressions (percepts); this is not the case for abstract concepts, which are however no less meaningful. In other words, every thought that is effable (capable of being expressed) can be regarded as a concept.

An abstract syntax and model-theoretic semantics for Conceptual Graphs are provided in [28]. Within the Sowa approach, concepts are related to symbols and to referents in a 'meaning triangle' depicted in Fig. 1 [27] and based on [29].

Here, a mental CONCEPT is symbolized by SYMBOL which could be a word. The extension of the SYMBOL is all the things it stands for – which could be thought of as instances of the CONCEPT. In other words, CONCEPT is an intentional abstraction of the common properties of the REFERENTs to which it refers. According to Sowa [30], every concept has a *type*, which defines a class of referents. The *type* is therefore a specification for a set or class of entities in some domain. The type specification is a monadic predicate; e.g., in logic, $P(v)$.

The meaning triangle can be given a mathematical interpretation based on first order model theory. To say that a symbol 'symbolizes' a concept means that SYMBOL gives CONCEPT an identity; e.g. a name or a tag. In the model theoretic semantics of Conceptual Graphs, CONCEPT has a type, which defines a class of REFERENTs. The association of SYMBOL with REFERENT in the triangle is an instance of what is called an *interpretation map* in the mathematical theory of models.

In a sentence expressed in a language (L) about one or more concepts (as in the conceptual graph in Fig. 1), each concept will have an associated meaning triangle. The sentence applies to every referent that belongs to the class; and to the symbol of the concept. The symbols in the sentence belong to what are collectively called the *signature* ($\sigma$) of the language (L).

A model of the sentence involves two things: (i) a choice of referents; and (ii) the interpretation of the signature in a way that results in the sentence being true (i.e. the interpretation satisfies the sentence). Note that the interpretation map is specified at the level of symbols rather than the level of the entire sentence.

Conceptual graphs are symbolic expressions that represent relations between concepts or their referents. The elementary conceptual graph as depicted in Fig.1 without any indication of directionality is intended to be read from left to right for ease of expression in natural language. In this case, 'Concept 1 has Relation to Concept 2'. Sowa uses the following notation:

$$[Concept\ 1] \rightarrow (Conceptual\ Relation\ 1 \rightarrow 2) \rightarrow [Concept\ 2] \quad (2)$$

This is interpreted as meaning that a concept [Concept 1] is related to another concept [Concept 2] via a conceptual relation (Conceptual Relation 1 → 2). Concepts and conceptual relations are further defined in terms of their types and any constraints applying to them. This is equivalent to having limited knowledge captured in a domain ontology.

To represent the conceptual relation in Fig. 1 using the language of the Predicate Calculus, each of the types associated with the concepts needs to define a set of referents. The model theoretic semantics for this graph are then stated in terms of the referents (entities), $\exists v_1, v_2: P(v_1, v_2)$ where $v_1$ and $v_2$ take on the values 'Entity 1' and 'Entity 2.' Each entity is a referent of the respective concept, and the predicate letter $P$ takes on the value 'Relation'. Higher order relations may be required for practical applications and are admissible in conceptual structures.

The above is an example of the 'principle' prescribed by Brooks. However, it will be seen in Section V that not every concept can be represented in the model theoretic semantics of a conceptual graph. This is due to fundamental limitations in the Predicate Calculus.

According to Sowa, the meaning of any concept is acquired solely through its contextual relations as part of a subjective 'semantic network', whose nodes include other concepts, percepts, motor mechanisms, emotions and other mental phenomena. In other words, what gives a conceptual graph meaning is its interpretation via a mind's semantic network. This echoes a subjective view of architecture described in Section II.

Some of the semantic network can be formalized in a way that makes its internal workings at least partly explicit. This 'explication of the implicit' can be thought of as the manifestation of concepts and conceptual relations (as brought together in conceptual graphs). Making concepts and their relations explicit allows logical reasoning to be applied – which is the basis of mathematics, science and engineering. The syntax and model theoretic semantics of conceptual graphs provide a first order language that can be used to implement the Essential Architecture Definition process to be introduced in Section IV.

### D. Analysis of Distinction between Key Terms

This section concludes with a consideration of everything presented in the paper up through Section III-C for the purpose of assessing the distinction between architecture and key terms such as: *property, concept, embodiment, element*, and *relation*. *Class* will also be addressed although it was not one of the key terms in the adopted definition.

It is important to keep in mind that the essential definition of architecture makes no direct reference to either of the terms *element* or *relation*. Nonetheless, the term *structure* as defined in Section I-B can be interpreted as a relational structure in model theory, which is in fact comprised of elements and relations. Therefore, these two terms will be considered together as (relational) structure. The essential definition also makes no reference to the term system; and hence, it has no inherent dependency on its definition.

For the essential definition proposed in this paper, architecture is a property but not every property is architectural. This is because type is a specialization of property; and architecture has been defined as structural type. Similarly, architecture is not a concept per se. When expressed in the syntax and model theoretic semantics of knowledge representation, every concept has a type. Concepts that have a structural type can be architectural. This is consistent with the adopted definition. It should also be noted that architecture is not conceptualization. Some of the 1990s definitions considered it to be; but it would be better to regard conceptualization as an activity in an architecture definition process in order to avoid confusion.

Furthermore, architecture is not a structure. Rather, it is a structural type. Because type is a specialization of property; and properties define a class or set but are distinct from what they define, architecture is not a class and is also not a conceptual structure. The distinction between models and architecture is then clear. Models are sentences interpreted into structures. (The sentences express theories or concepts.) Consequently, if architecture is a model; it is then a structure, which as previously noted is not the case.

Finally, *architecture is not embodiment into elements and relations*. The essential definition associates structural type with architectural properties that can be implemented by that type. In some respects, this is similar to the idea of 'embodiment'. However, the embodiment of a property (or concept) into elements and relations (i.e. a relational structure) would be an architectural model (in the sense of Tarski). Thus, the use of the term embodiment in the adopted definition, including the refinement in [16] and [17] mentioned in Section II-B, risks confusion between architecture and model. This can be an issue in practice with the adopted definition for architecture description.

In summary, models are interpretations of sentences (e.g. theories or concepts) into structures. Architecture defines the type of structure to be used as well as further properties that must be satisfied (e.g. constraints and other relations); but is not itself the structure. An architectural class is an implementation of the type and properties. Given an object of interest, architecture is then the basis for constructing models that describe or define the object. The mathematical foundation of the essential definition provides distinction between key terms related to architecture.

### IV. ARCHITECTURE DEFINITION AND SYSTEM DESIGN

As formalized in this section, the Essential Architecture Definition process involves: (i) defining theories and concepts relevant to an interrelational structure, (ii) specifying the symbols for the language associated with the theories, and (iii) specifying fully interpreted models of the theories. Using this process, the precepts of MDA can be extended to engineering in a straightforward way for general system design and not just software development. At this level of abstraction, the process can be applied broadly to the life-cycle specification of a system.

*A. Architecture Definition and MDA*

Essential Architecture Definition is at a higher level of abstraction than the processes in currently adopted standards. Based on the first order model theory of the previous section, it can be understood as follows in a mathematically rigorous way:

i. Define a sequence of theories $\{T_j\}_{j=1,\ldots,n}$ relevant to an interrelational structure $(A = R_0, R_1, \ldots, R_n)$ described in natural or technical language relevant to a conceptual graph that formalizes a concept.

ii. Specify a sequence of signatures $\{\sigma_j\}_{j=1,\ldots,n}$ for the language(s) associated with the theories.

iii. Specify fully interpreted first order models using the interpretation of the theories $\{T_j\}_{j=1,\ldots,n}$ in terms of the relational structures $\{(R_{j-1}, R_j)\}_{j=1,\ldots,n}$.

In model theory, a *theory* is a consistent set of sentences in a language appropriate for the types of structure considered [25].

When an OMG MDA PIM is implemented for a software system using the Unified Modeling Language (UML), the functionality of the system can be specified by means of Use Case diagrams and behavior of the system through Activity diagrams. The specification of functionality in a use case is an example of pairing a property with a structural type that can be implemented by a class. The use cases can also be represented in terms of conceptual structures. Functionality is therefore implemented in two classes of structure of that same type. Each class will be useful for engineering but in different ways. One will be a class of conceptual structure; and the other will be an object-oriented class (a use case) that can be used in the software PIM. One type of sentence relevant to a use case is a statement of functional requirements. The quantitative constraints to which the system must comply will be captured through interpretation of the conceptual structure. The use of *common sentences* for functional requirements interpreted into these structures (which are of the same type) will result in equivalent models and assure architecture concordance. The sentences are *theories*. When they refer to a system, they are concepts about the system.

The use cases will form a relational structure that will have interrelations when the system is fully specified. The general form of the type of use case for representing functionality is a verb – noun phrase which associates an actor that interacts with the system by means of the use case. Because the verb establishes a relation between the system and actor, the phrase is a monadic predicate and therefore is a structural type. This type of use case can be represented as an interpretation of the sentence, 'system interacts with actor'. The verb phrase is 'interacts with' and the actor is the noun. The structure represented by a Use Case diagram is a *partially interpreted model* of the sentence. In this functional representation of the system, elements of the system are not yet specified; rather only functional relations are.

A so-called graphical model such as a Use Case diagram represents a mathematical relational structure of vertices and edges [31] into which sentences have been interpreted. The vertices are sometimes called the nodes of the graph. The nodes and edges are *symbols in the signature* of the model-based language (e.g. UML) in which the theories have been expressed. Typical symbols for the system (with its boundary) and the actor are a rectangle (which contains the name of the system and the use case(s)) and a stick figure (which is given the name of the actor). These are the nodes. The edge is a line that represents an (interaction) association between the system and actor.

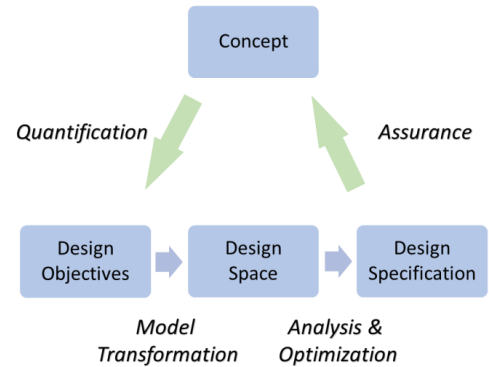

Fig. 2 Model Driven System Design

A conceptual graph on the other hand depicts a relation among entities. Each node in a conceptual graph is either an entity or a relation. The edges in the graph serve only to associate the nodes but otherwise have no meaning. The conceptual graph is an interpretation of a sentence into a relational structure; hence it is a representation of a model.

Therefore, graphical models of functionality represent models of a functional architecture for a system. This application of Essential Architecture Definition is in agreement with what is meant by an MDA PIM. It is also an example of how it can be applied broadly to a life-cycle specification of a system.

*B. Extension of MDA to Model Driven System Design*

Essential Architecture Definition supports an extension of MDA to design and development that encompasses both hardware and software. The nature of engineering considered in [9] is based on its evolution in Western civilization [32]. Engineering of a product begins with craft but must be honed using the laws of science for the precision and repeatability needed in commercial production. The following definition aligns well with the essential definitions that have been proposed:

*Engineering* is a practice of concept realization in which relations between structure and functionality are modeled using the laws of science for the purpose of solving a problem or exploiting an opportunity.

The architectural viewpoint in this definition is worth noting. The association of structure and property (i.e. functionality) conforms to the pairing in the essential definition of architecture. Also, rather than focus on functionality and behavior, as MDA does with the PIM; the emphasis is on structure and functionality. In the higher level of abstraction of the essential definitions; behavior is a type of structure and functionality is an architectural property. As with other terms defined in this paper, there are numerous definitions of engineering in multiple relevant standards; e.g. [14] and [33].

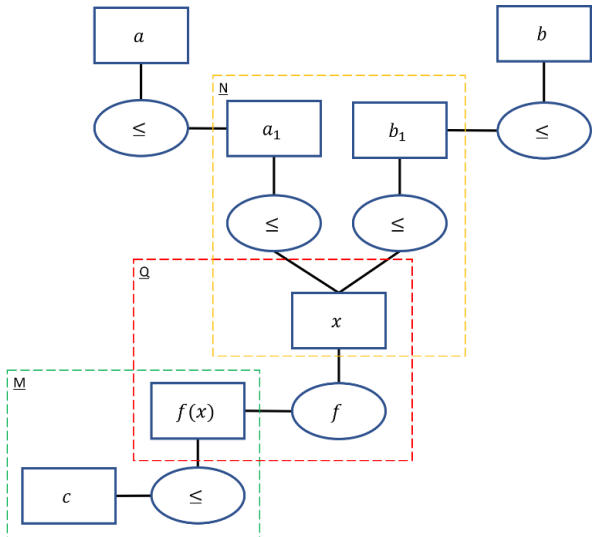

Fig. 3. Conceptual Graph of a Constraint Model Transformation.

In this definition of engineering, relations between the structure and functionality of a system are the basis for concept realization. The centrality of the term 'concept' to both architecture and engineering makes possible the definition of a process for system development that mirrors the underlying process of MDA for software development. This will be termed as Model Driven System Design (MDSD) and the process is depicted in Fig. 2. It is based on four types of models: (i) Concept, (ii) Design Objectives, (iii) Design Space, and (iv) Design Specification. As depicted in the figure, these models are linked by Quantification, Model Transformation, Analysis and Optimization, and Assurance. MDSD is not intended to be an end-to-end life-cycle process for system development.

Requirements can be regarded as concepts about the system and also linked with system architecture. In this way, requirements and other concepts about the system can be used as a starting point for MDSD. It should be noted that Requirements Definition is one of the two most used technical processes in the practice of MBSE. Architecture Definition is the other [34].

In modern software and engineering practice, the architecture of a system is generally considered to be the first artifact of design [21]. For MDSD, the Essential Architecture Definition process can be applied to but is not limited to Concept Definition and specification of Design Space Architecture, the models of which become the subject of analysis and optimization for Design Specification. Thus, the outcomes of architecture definition as conceived in this paper can include the early artifacts of system design but are not limited to these.

### C. Constraint Driven Design

The engineering and design of complex systems can be driven by constraints on design objectives such as stakeholder requirements or compliance to regulations. This type of constraint can impose limitations on the design space. Constraint Driven Design (CDD) is a system design approach in which design objective constraints are a central concern. A solution obtained from such a design approach aims to satisfy all design objective constraints [35].

CDD conforms to MDSD in a mathematically natural way. When the system requirements (to include design objective constraints) are specified using conceptual structures; the quantification of concepts into design objectives can result in mathematical relations. For a given design objective $z$, the simplest type of constraint is in the form $z \leq c$. This could be, for example, a mathematical interpretation of the concept "the measurement $z$ of a specified emission must be less than or equal to a regulatory constraint $c$."

The design space typically comprises several design variables (which are sometimes referred to as factors). For each variable $x$, constraints of the form $a \leq x \leq b$ are imposed. This can be based on domain knowledge, for example. Each constrained one-dimensional factor of the design space will be in the form of such an interval. As a concept, design variable constraints are represented graphically in Fig. 3. The bounded regions illustrate the structure of the ROSETTA matrix representation depicted in Fig. 9 explained in Section V-C.

In a two-dimensional design space, constraints of this type on each factor will be in the form of a rectangle, which is referred to as a feasible region. A two- or higher dimensional 'rectangle' formed by constraint relations on each variable is called an *orthotope* in multi-dimensional Euclidean space. If these are the only constraints on the design space, the design specification permits independent choices of values for each design variable (within the orthotope). Design decisions can be made 'one factor at a time' because of the independence.

The feasible region though is usually constrained by further relations. Joint constraints on the objectives are transformed into further relations on the design variables that can change the geometry of the feasible region in ways that can complicate the specification of individual design variables. The transformed relations can arise from relations between design objectives and design variables; for example, a response surface $f : X \rightarrow Z$ that might be based on models from engineering or science. However, the (joint) specification of continuous design variables should always be in the form of an orthotope.

The CDD model transformation derives from the mapping $f^{-1} : P(Z) \rightarrow P(X)$ where $P(\cdot)$ denotes power set (i.e. the set of all subsets). For a one-dimensional response surface, a constraint given by $f(x) \leq c$ (for a given $x$ in the design space) implies a condition on the design specification: $a_i \leq x \leq b_i$ for each individual design factor. Fig. 3 is a conceptual graph of such a model transformation. CDD implements MDSD by specifying transformations like this. Fig. 4 depicts how this can affect the geometry of the design space for two key design variables in the emissions case study in Section V.

In general, every orthotope that satisfies the transformation of objective constraints into the feasible region is therefore a candidate solution (design specification) for the associated CDD problem. This family of orthotopes forms a relational structure on the constrained feasible region determined by the CDD transformation. This defines a *structural type* that can be called *constraint driven structure (of the design space)*.

The processes of interpreting concepts into design objectives followed by CDD model transformation into the design space are examples of technical processes for Essential Architecture Definition as given in Section IV-A.

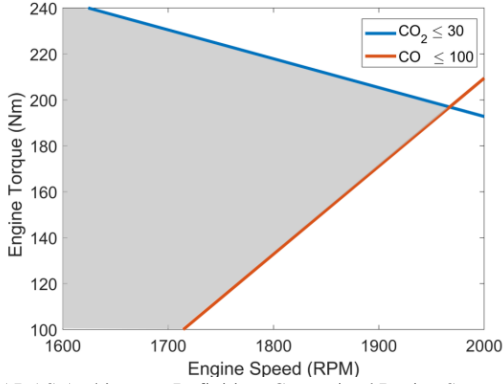

Fig. 4. ADAS Architecture Definition: Constrained Design Space.

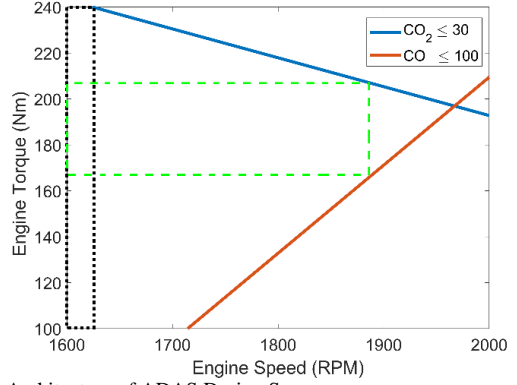

Fig. 5. Architecture of ADAS Design Space.

*D. Application to ADAS with Emission Control*

A coordinated control architecture design [36] for a class of emissions control problems will now be investigated to demonstrate MDSD at the simplest possible level. A design space for an advanced driver assistance system (ADAS) will be architected to integrate an emissions control function with an existing cruise control system. A modular system architecture will be assumed for which the integrated solution does not require redesign of the existing system. Essential Architecture Definition will be used to define an architecture for the Design Space.

Two design objective constraints will be specified for emissions and two design variables for the engine. In this simplified problem, the two objectives are $z_1 = CO_2$ emission and $z_2 = CO$ emission, measured in g/km. The constraints are $z_1 \leq 30$ and $z_2 \leq 100$. The design variables will be engine speed, $x_1$ in revolutions per minute (RPM) and engine torque, $x_2$ in Newton-meters (Nm). An engine map has been measured for the state space $1600 \leq x_1 \leq 2000$ and $100 \leq x_2 \leq 240$. As a mathematical mapping, the map is defined by two response surfaces $f(x_1, x_2) = z_1$ and $g(x_1, x_2) = z_2$. In practice, the map is normally a table of discrete measurements.

The outcome of CDD model transformation is the shaded region of the design space displayed in Fig. 4 by the transformation of the constrained objectives into engine torque and engine speed through the inverse mappings of the two response surfaces. The ADAS cruise-control functionality includes providing torque demands to the engine to: (i) maintain a constant vehicle speed (which in a fixed gear corresponds to a constant engine speed); and (ii) accelerate the vehicle (which in a fixed gear corresponds to increasing the engine speed).

The integration strategy for the ADAS emissions control functionality is to mediate torque demanded by the driver before it is communicated to the existing cruise control system. The demanded torque presented to the system may need to be reduced in order to ensure the emissions constraints are not violated.

At the start of either a constant speed cruise or an acceleration action, the engine is initially at a given state (RPM, torque) within the shaded region in Fig. 4. The simplest initial starting point would be towards the center of the region. When the demand for increased torque is sent to the engine, the outcome will be increased RPM. If the engine state starts from near the center, then there is no need to manage the relation between torque states and engine speed (for the purpose of emissions control) until the state approaches one of the two boundaries. Keeping the state within a rectangular region (a two-dimensional orthotope) in the constrained design space is a simple and easy method to integrate ADAS emissions control functionality with the control system.

Any rectangle that fits within the shaded region is in principle suitable for specifying a region of control, i.e. a set of combinations of engine torque values and speeds that comply with the constrained design space. Rectangular regions such as these are orthotopes (two dimensional in this case). They are the basis for an interrelational structure on the constrained design space as well as design specification, i.e. a relational structure of orthotopes interrelated by shared relations with the constraints.

There are infinitely many orthotopes, two of which are depicted in Fig. 5. The 'tall narrow' orthotope reflects allowing for maximum variation in torque but only in a narrow range of engine speeds. The restriction in engine speeds (which is seen by the narrowness of the orthotope) would be appropriate for the function of maintaining a constant vehicle speed (corresponding to engine speeds near 1600 RPM). The mediation of torque demands should be minimal in this case even over hills, as the existing cruise control system will manage the engine speed to maintain a constant vehicle speed. If the initial engine state is somewhat greater than 1600 RPM, a different orthotope will need to be selected towards the right of the one in the figure. This will necessarily reduce the vertical extent of the orthotope, and consequently the range of torque demands permitted.

A more 'balanced' orthotope is also shown in the figure; but it clearly reduces the range of torque outputs permitted. This orthotope could also be useful for the function of accelerating because the increase in engine speed for much of the region is not at risk of taking the engine state past the constraints on emissions.

Note that the two orthotopes in Fig. 5 enjoy the property of *maximality* within the constraint driven structure. This means that neither of them can be contained by another orthotope that conforms to the constraints. In other words, an orthotope is maximal if any other orthotope that contains it must violate at least one of the constraints on the design space.

The result is an interrelational structure formed of 'maximal' orthotopes. Identifying this set of orthotopes as a desirable (if not required) feature of candidate design solutions marks a completion of Essential Architecture Definition in the MDSD process for this example. In engineering practice, specifying architecture for the design space can therefore help to point the designer as to where to find suitable solutions.

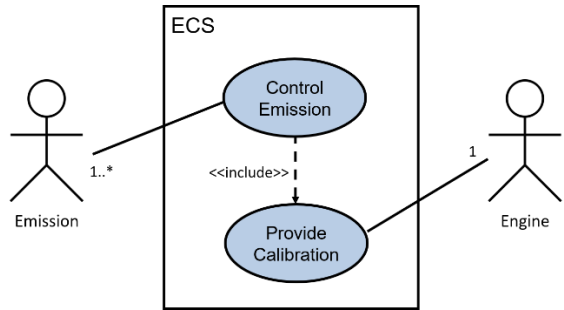

Fig. 6. Use Case Diagram for the ECS

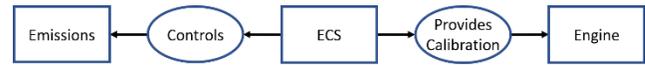

Fig. 7. Conceptual Graph for ECS Functionality

This design of ADAS illustrates how the process of Essential Architecture Definition can guide the designer in the selection and specification of models based on structural types and architectural properties. The practical nature of the example also illustrates how the definition process for a system need not be a linear process. The designer is free to think in terms of models and later consider what types of structure they have been expressed in. For conceptual integrity and architectural concordance in design iterations though, structural types and architectural properties should be specified as early as possible.

Architecture definition as demonstrated in this example is seen as a general structuring process that provides a basis for modeling (among other things). The interplay between architecture definition and the specification of models is central to engineering practice as defined in this section. The structural types that have been used are inherently simple but require insight, if not deep understanding of the system. Specifying the right types of structure is therefore critical to system design.

## V. EMISSIONS REDUCTION CASE STUDY

The case study will develop a high level design for an emissions control system (ECS) that will demonstrate the engineering utility of the definitions and processes offered in this paper. The design will feature but be limited to a high level object-oriented PIM level specification for the ECS software and a calibration method that can be used to reduce emissions. The object-oriented models for the MDA PIM will be linked to engineering models for engine calibration using conceptual structures as in Section IV. As previously noted, this provides conceptual integrity and architectural concordance (in the ECS). The case study will then introduce recent advancements in commercial quality calibration methods [37] by building on the design space architecture for the ADAS in Section IV. Algorithms can be implemented using a relational-oriented mathematical framework that can accommodate structures associated with maximal orthotopes in the design space. The case study concludes with a consideration of how the definitions and processes also can be part of systems engineering standards.

### A. Engine Calibration Problem Definition

In the case study, the problem to be solved is that emissions from automotive diesel engines must not exceed regulatory constraints when the vehicle is driven in a variety of profiles such as those specified in [38]. The primary source of emissions originates from inefficient combustion in the engine that leads to undesired chemical products such as nitrogen oxides ($NO_X$). In modern diesel engine cars, emissions are often reduced using aftertreatment technologies such as Diesel Particulate Filters and Lean $NO_X$-Traps.

One approach to improve the efficiency of the chemical reactions that occur within the engine combustion chambers is to design an optimal engine map. In order to achieve current regulatory constraints, a lower-level engine map is used that involves tuning of the engine calibration parameters, such as Mass Air Flow (MAF), Fuel Rail Pressure (FRP), and Low Pressure Exhaust Gas Recirculation (LP-EGR).

The ECS controls emissions (indirectly) through the provision of calibration data to the engine. This is the basis of the functionality of the ECS. When the ECS is considered as a system in this case study, it will have direct interactions with the engine and an ADAS which mediates commands from the driver to the engine, and which informs the ECS of the commands sent. The system elements will be an electronic control unit with associated software applications and an engine map. The functionality and behavior of the ECS will be the subject of Section V-B. The association of calibration methods for managing the engine map will be the subject of Section V-C.

### B. Specification of High-level MDA PIM

MDA development focuses first on the functionality and behavior of a system by means of a platform-independent model (PIM) that separates application logic from underlying platform technology. A use case diagram will be specified for system functionality and an activity diagram for system behavior. For simplicity, different levels of abstraction will be used: the ADAS will be included in the activity diagram but not the use case diagram. Also, the emissions will not be included in the activity diagram. (The diagrams are an output of architecture definition as previously discussed.) Software implementation and PSM level details of this elementary software architecture for the ECS will not be pursued. These are routine in MDA. The modeling implications for system architecture in the case study are of greater interest than the detailed modeling of software.

Each of the two use cases in Fig. 6 depicts something that the ECS system is intended to do: (i) control emissions (from the engine), and (ii) provide calibrations (to the engine). Each is an 'interaction' relation between the system and an 'actor' (in this example, the engine or its emissions). An inclusion relation between the two use cases is used to model the indirect effect of the ECS on emissions. As in Section IV, the use cases depict a relational structure comprising two relations that are an interpretation of system functionality.

As in Section IV-A, two types of graphical models can be constructed that interpret the sentences associated with the functional requirements. The result is depicted in Figs. 6 and 7. The conceptual graph in Fig. 7 (which is comprised of two relations joined by the ECS) also represents a relational structure comprising the two relations that are an interpretation of system functionality. This view of system functionality is external to the ECS (i.e., interactions with its environment). Internal details of the system are suppressed.

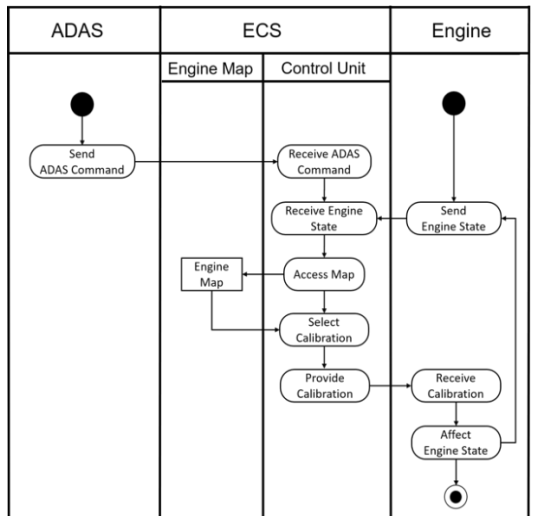

Fig. 8. ECS Activity Diagram

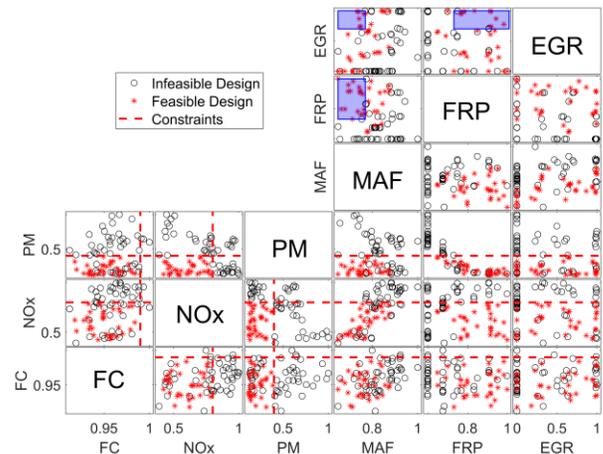

Fig. 9. ROSETTA framework for emissions control

However, when the two figures are closely compared, it is apparent that Fig. 6 has a relation of *inclusion* between the two use cases, which cannot be captured as a relation in the conceptual graph in Fig. 7. The syntax of a conceptual graph must conform to the Predicate Calculus which only permits relations between entities (e.g. concepts) and not relations between relations. Inclusion is an example of interrelationship.

The *functional structure* of the ECS is therefore represented in Fig. 6 by two relations (use cases) and one interrelation (inclusion). Note then that interaction is sufficient to define *system functionality* but not sufficient to define the *functional structure* of the system (when the internal details are considered).

The structure represented in Fig. 6 is an example of an object-oriented structure that can be used to model the interrelations of the *system*. The interrelational structure creates an association between *system* and *architecture* (as noted in the Introduction). Systems endowed with a structure of this class can be said to conform to the functional architecture of the ECS.

It should also be noted that the use case relations depicted in the conceptual graph (Fig. 7) and the use case diagram (Fig. 6) are semantic predicates. As depicted, the statements in the figures are undecidable. They are sentences that have partially interpreted models in the sense of Tarski theory. There is not yet enough information to determine if the predicate sentence (e.g. a well-formed formula quantified by an existential operator) is true or false (*in the domain*). The engineering problem is to specify the system (model) at a level of detail from which it can be realized in conformance to the functional architecture as well as other defined requirements. This will validate the fully interpreted model; and is concept realization.

An Activity Diagram must also be given to complete the specification of the PIM. The influence of the ECS on the emissions from the engine is achieved by its use of the Engine Map and a provision of calibrations in real time to the engine. The diagram in Fig. 8 represents the high level behavior of the ECS in its relation to the ADAS and engine. It depicts a structure that has been realized through structures of behavioral type. This type of structure could also be represented in a conceptual graph; but this will not be needed in this case study.

Use Case and Activity Diagrams represent the functionality and behavior in the PIM for the software architecture of the ECS. Specifying a calibration method will complete the design.

### C. Structures for Calibration Method Design

In this and the next section, an integration of the Relational Oriented Systems Engineering Technology Tradeoff and Analysis (ROSETTA) framework and Axiomatic Design [39] will be used to show how a calibration method can be developed to seek robust calibrations for emissions control. In a modern automotive vehicle, dozens of calibration factors must be considered; but in practice, a dozen or fewer dominant factors are typically sought. These will become the design variables in an MDSD problem. As with the ADAS example, maximal orthotopes for these variables will be identified within a constrained design space. For simplification, the problem will be reduced to three design objectives and three design variables that are joined together by three response surfaces.

A matrix formulation of ROSETTA provides a mathematical implementation of MDSD that is a framework similar to the Quality Function Deployment (QFD) House of Quality [40]; but can be used to translate expert opinion into mathematical relations. For this reason, the name of the framework bears an intentional similarity to the Rosetta stone which provided the means to interpret between the Greek, Egyptian and Hieroglyphics demotic languages. In an Essential Architecture Definition process for CDD, ROSETTA provides a facility for interpretation of models between systems engineering and systems (and software) architecting. It supports rigorous development of structures for system design as in the example in Section IV-D. ROSETTA will be used to extend constrained design spaces such as in Fig. 4 to higher dimensional spaces.

A mathematical implementation of MDSD using ROSETTA is depicted in Fig. 9. The dots are normalized data from a design of experiments. The data are structured using binary relations depicted in the $1 \times 1$ matrices that correspond to pairings of the design objectives and design variables. (These matrices represent a structural type of binary relationship.) Outcomes of model transformations like the one in the conceptual graph of Fig. 3 are depicted as regions bounded by the dashed lines (that correspond to emission levels constrained by regulations). The feasible design points are those that comply with the constraints. Fig. 9 is a multi-variate expression of Fig. 3.

TABLE I
RESPONSE SURFACE COEFFICIENTS

|  | $\beta_0$ | $\beta_1$ | $B_2$ | $B_3$ | $\beta_{11}$ | $B_{22}$ | $B_{33}$ |
|---|---|---|---|---|---|---|---|
| $CO_2$ | 5.97 | -1.21 | -11.31 | -0.07 | 0.30 | 6.27 | 0.03 |
| NOx | -4.01 | 6.53 | 2.89 | -0.24 | -2.37 | -1.72 | 0.03 |
| Soot | 1.22 | -0.34 | -0.42 | -0.02 | 0.27 | 0.27 | 0.03 |

The aggregation of the pairings of the three design objectives (FC, NOx, and PM) with each other form a 3 × 3 structure that is referred to as the M-matrix. The cells contain the constrained binary relations of the requirements. This is the objectives matrix. Similarly, the design matrix is referred to as N and is formed by the aggregation of the pairings of the design variables (MAF, FRP, and EGR). The Q-matrix depicts the relationships between the design objectives and the design variables. It corresponds to a transformation of the design objectives into the design variables that will further constrain design solutions. The feasible designs are thus transformed into the N-matrix but the simplicity of the constraint lines in the M-matrix is lost. The two dimensional projections of the maximal orthotope in Fig. 10 into the ROSETTA framework (i.e. the shaded rectangles in the cells) make it possible to visualize higher dimensional solutions.

Relations in the Q-matrix can be mathematically approximated by a set of response surfaces given by $z_i = f_i(x_1, x_2, x_3)$. Each matrix element in the Q-matrix also corresponds to the sensitivity between a pair $z_i$ and $x_j$. The gradient line slopes are calculated by taking the partial derivative, $(\partial z_i)/(\partial x_j)$, of the response surface at the initial design point. In this specific example, the response surfaces are approximated by pure-quadratic functions of the form:

$$z_i = \beta_0 + \sum_{j=1}^{N} \beta_j x_j + \sum_{j=1}^{N} \beta_{jj}(x_j)^2 \quad (3)$$

The sets of normalized coefficients, $\beta$, as presented in Table I, have been derived from an engine test bed for an engine speed of 1350 RPM and a torque demand of 25 Nm. As such, the slopes of the local gradients are determined by the expression:

$$\left.\frac{\partial z_i}{\partial x_j}\right|_{x_j=x_{j0}} = \beta_j + 2\beta_{jj}x_{j0} \quad (4)$$

where the initial design value for $x_j$ is denoted by $x_{j0}$

In higher dimensional multiobjective multiattribute problems, the approximation by response surfaces in the matrix representation provides a method to reduce complexity in the system analysis. Therefore, ROSETTA can be used both to structure information and to manage complexity.

### D. Specification of a Calibration Method

The use of maximal orthotopes, as in Fig. 5, has been extended to the emissions problem. ROSETTA and the Nam

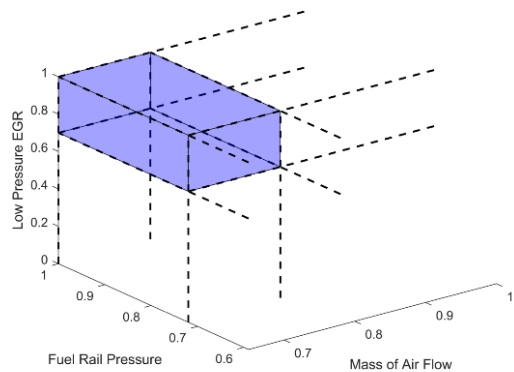

Fig. 10. A Maximal Orthotope for Emissions Control

Suh Axiomatic Design theory have been integrated to develop algorithms to seek maximal orthotopes of calibrations derived from an engine map to be provided to the engine in real time, as in the Activity Diagram depicted in Fig. 8.

The result is a recent advancement in commercial quality calibration methods that has been published in a patent with Jaguar Land Rover [37] to which the reader is referred for details of the algorithm. The essential idea is to rank the variables then create a maximal orthotope one factor at a time by pushing each factor to its constraint limit in the order of the ranking. Because of the interrelationships between the design variables (i.e. the calibration factors), every step of the process reduces the available design space. Fig. 10 depicts a maximal orthotope associated with the framework in Fig. 9.

### E. Model-based Systems Engineering Standard

The concepts of CDD are therefore seen to have substantive engineering utility in the context of MBSE based on relational structures formed of constraints on design objectives that are transformed into structures in the design space. The engineering utility of the essential definitions and the architecture definition process offered in this paper have been clearly demonstrated. Standardization of terms and concepts in MBSE can also benefit from the definitions, architecture definition process, and mathematical interpretations put forth in this paper.

A standard that provides evidence of this claim is the OMG UPR: the UML Profile for ROSETTA [35]. The profile specifies a comprehensive facility to structure information in support of model-based analysis for architecture optimization and system design. The facility can also support efficient and effective transformation of the system architecture into detailed system design. Such a facility could be used, for example, to capture and structure information from a system model for a complex system like that of the calibration problem for emissions reduction.

UPR supports information structuring for the *quantification* and *model transformation* portions of MDSD depicted in Fig. 2 but was not intended to define new techniques for model-based analysis and optimization. The implementation of CDD through ROSETTA, on the other hand, is intended as a new technique. Information captured through a UPR facility can be used to populate the framework in a way that maintains conceptual integrity of the various models of interest.

UPR comprises five normative packages: (i) Foundations, (ii) Operators, (iii) Design Objective Constraints, (iv) Design Variable Constraints, and (v) Relational Structure. Operator and Constraint stereotypes provide interpretations into UML in each of the binary relations in Fig. 3 for CDD model transformation. The mapping $f(x)$ can also be interpreted when it is in the form of a polynomial response surface as in Fig. 9 and Table I.

Therefore, conceptual graphs as in Fig. 3, matrix arrays as in Fig. 9, and the UPR diagrams (that implement its stereotypes) can be used to implement models in structures of constraint driven type as introduced in Section IV-C. The essential definitions for structure and architecture together with the explanation of Tarski model theory and conceptual structures in Section III offer a mathematical basis for UPR.

## VI. Conclusions and Future Work

New intuitive fundamental definitions complemented by mathematical interpretation have been contributed and used to specify a mathematically based technical process for architecture definition that can be applied to the life-cycle specification of a system. Essential Architecture Definition is a general structuring process that involves: (i) defining theories, (ii) specifying symbols for the language associated with the theories, and (iii) specifying fully interpreted models of the theories. Its validity has been demonstrated through several elementary examples in the practice of system and software design. Engineering utility has been demonstrated in a detailed case study on recent advancements in commercial quality calibration methods for diesel engine emissions and by effective modeling of constraint driven design (CDD) problems for which the Object Management Group (OMG) has recently adopted an open international standard (UPR: the UML Profile for ROSETTA).

Ambiguities in the adopted definition of architecture have been clarified through mathematical interpretation. As defined in this paper, *architecture* is consistent with the adopted definition but is at a higher level of abstraction. Models are interpretations of concepts (theories) into structures. Architecture specifies structural type and properties to be implemented.

The essential definitions, the Essential Architecture Definition process, and advances put forward in this paper will be carried forward along lines of both research and commercialization. The advances are being applied to a facility and methods for CDD, and to the electronic architecture of vehicles [41]. The research will use Category Theory as an advanced structuring method for architecture definition. New formalisms can enhance systems engineering tool development and evolve the scientific basis of the discipline.

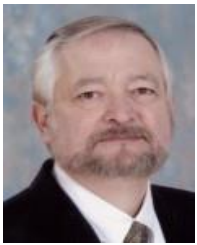

**CHARLES E. DICKERSON** (M'06–SM'17) received the PhD in mathematics from Purdue University, West Lafayette, Indiana, USA in 1980.

He is Professor and Chair of Systems Engineering at Loughborough University, UK. Previously he was Technical Fellow in BAE Systems in the USA. His aerospace experience further includes the Lockheed Skunkworks and Northrop Advanced Systems. He was also a member of the Research Staff at Lincoln Laboratory, Massachusetts Institute of Technology. He has served secondments as the Aegis Systems Engineer for US Navy Ballistic Missile Defense; and Director of Architecture for the Navy's Chief Engineer. He is currently Co-chair of Mathematical Formalisms in the Object Management Group (OMG).

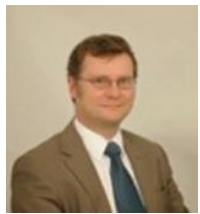

**MICHAEL WILKINSON** received the PhD in theoretical physics from King's College, London, UK in 1985. Previously, he received the BSc in physics from the same institution.

He is Chief Technologist at BAE Systems Maritime, Barrow-in-Furness, UK, and is a visiting Professor at Loughborough University. Previously he was a Technical Director and Professional Head of Discipline for Systems Engineering at Atkins. Prior to that he was Technical Director of the Niteworks partnership of the MOD. He has served as President and Academic Director of the UK Chapter of the International Council on Systems Engineering. He is currently Co-Chair of the INCOSE UK Chapter Architecture Working Group.

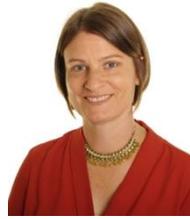

**EUGENIE HUNSICKER** received the PhD in mathematics from the University of Chicago, Chicago, Illinois, USA in 1999. She previously received the BA in mathematics from Haverford College, Pennsylvania, USA in 1992.

She is currently a Senior Lecturer in Mathematics at Loughborough University, UK. She was previously employed at Lawrence University in Wisconsin, USA. Her research is in analysis and topology, as well as applied statistics, especially for high dimensional and non-Euclidean data. Dr. Hunsicker is a member of the London Mathematical Society and the Royal Statistical Society.

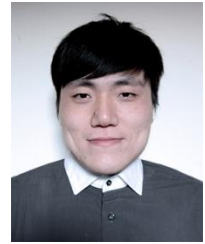

**SIYUAN JI** (M'16) received the PhD in physics from University of Nottingham, Nottingham, UK in 2015. He previously received the MSc degree in physics from the same institution.

He is currently a Lecturer in the Department of Computer Science at the University of York, UK. Previously he was a Research Associate at Loughborough University. His research is focused on systems engineering methodologies, model-based system safety and reliability assessments, product line engineering, and constraint driven design algorithms.

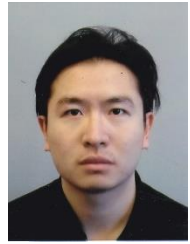

**MOLE LI** (M'19) received the bachelor's degree in computer science from the University of Hull in 2012 and the MSc degree from Loughborough University in 2013, where he is currently finishing the PhD.

He is a technical specialist in Rolls-Royce Control Systems in Derby supporting model-based systems engineering (MBSE) standardisation and developing methodologies and tools for whole engine life-cycle development. His research is in the integration of software product line engineering with MBSE.

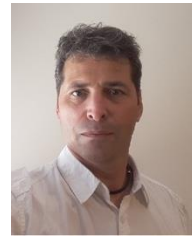

**YVES BERNARD** received the master's degree in biological oceanography from Pierre et Marie Curie Paris VI University, Paris, France in 1989.

He is MBSE support for the Process, Method and Tools Department of Airbus Defence & Space; and in the Airbus Avionics Department he is responsible for development of model-based methods applicable to critical embedded systems including both software and hardware components. He has been involved in a number of OMG standards to include UML 2.4 - 2.6, Precise Semantics of UML Composite Structures and State Machines (PCSC 1.0 and PSSM 1.0), Modeling and Analysis of Real Time Embedded Systems (MARTE 1.1), and UPR 1.0. He is Co-chair of the Revision Task Force for the Systems Modeling Language (SysML) and a member of the SysML 2.0 team.

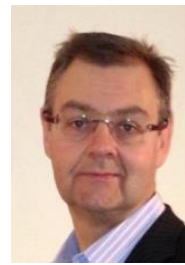

**GRAHAM BLEAKLEY** received the PhD in model-based systems engineering and design from City University, London, UK in 2001. He previously received the BEng in mechanical engineering from Southbank University, London, UK in 1992.

Dr. Bleakley is a Co-chair and lead architect of the OMG Unified Architecture Framework, which unifies DoDAF, MODAF and NAF into a single metamodel and commercially available framework for enterprise architecture and systems of systems development. He is also the IBM representative for SysML 2.0 at OMG.

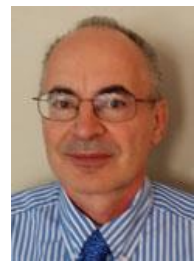

**PETER DENNO** received the master's degree in supply chain management from Pennsylvania State University and BA in mathematics from the University of Connecticut. He is completing a PhD in systems and industrial engineering at Loughborough University.

He is a computer scientist in the Engineering Laboratory, National Institute of Standards and Technology (NIST), USA, where he has led projects to facilitate the use of analytical methods in manufacturing. Previously he was in manufacturing and systems engineering at Pratt & Whitney.